\documentclass[journal=jctc,manuscript=article]{achemso}

\usepackage{mathtools}
\usepackage{caption}
\usepackage{subcaption}
\usepackage{algorithm}
\usepackage{hyperref}
\usepackage{xcolor}
\usepackage{algpseudocode}
\usepackage{tikz}

\usepackage[version=3]{mhchem} 

\author{Davide Castaldo}
\affiliation{Università degli studi di Padova, Dipartimento di Scienze Chimiche, Via Marzolo 1 - 35131 Padova (Italy)}
\author{Soran Jahangiri}
\affiliation{Xanadu, Toronto, ON M5G 2C8, Canada}
\author{Alain Delgado}
\affiliation{Xanadu, Toronto, ON M5G 2C8, Canada}
\email{alaindelgado@xanadu.ai}
\author{Stefano Corni}
\affiliation{Università degli studi di Padova, Dipartimento di Scienze Chimiche, Via Marzolo 1 - 35131 Padova (Italy)}
\email{stefano.corni@unipd.it}

\altaffiliation
{Padua Quantum Technologies Research Center, Università di Padova}
\alsoaffiliation[Second University]
{Istituto Nanoscienze—CNR, via Campi 213/A, 41125 Modena (Italy)}

\title[An \textsf{achemso} demo]
{Quantum simulation of molecules in solution}
\abbreviations{IR,NMR,UV}
\keywords{American Chemical Society, \LaTeX}

\begin{document}


\begin{abstract}
Quantum chemical calculations on quantum computers have been  focused mostly on simulating molecules in gas-phase. Molecules in liquid solution are however most relevant for Chemistry. Continuum solvation models represent a good compromise between computational affordability and accuracy in describing solvation effects within a quantum chemical description of solute molecules.  In this work we extend the Variational Quantum Eigensolver to simulate solvated systems using the Polarizable Continuum Model. To account for the state dependent solute-solvent interaction we generalize the Variational Quantum Eigensolver algorithm to treat non-linear molecular Hamiltonians. We show that including solvation effects does not impact the algorithmic efficiency. Numerical results of noiseless simulations for molecular systems with up to twelve spin-orbitals (qubits) are presented. Furthermore, calculations performed on a simulated noisy quantum hardware (IBM Q Mumbai) yield computed solvation free energies in fair agreement with the classical calculations. 

\end{abstract}

\section{Introduction}

Nowadays, multiscale modelling is a workhorse of computational chemistry and physics \cite{weinan2011principles,mennucci2019multiscale,segatta2019quantum, franco2019boosting}. Its recent development has been fueled by the constant quest to understand and harness more complex phenomena which necessarily call for the inclusion of details arising from the composite nature of the studied systems \cite{mennucci2019multiscale}. Such a demand for greater detail in the simulation of experiments is accompanied by an increase of the computational resources required. In particular, the need for more accurate molecular simulations using wave function based methods has motivated significant research efforts at the intersection of quantum chemistry and quantum computing. \cite{arguello2019analogue, nielsen2002quantum, deutsch2020harnessing, georgescu2014quantum}. 

In this work, we aim to contribute to both fields by reporting the first example of a hybrid quantum-classical algorithm (i.e., with the meaning given in the quantum computing literature to hybrid\cite{cerezo2021variational}, a protocol whose implementation relies both on classical and quantum computation) in which the quantum simulation of a molecular system takes also into account the presence of a solvating environment. The importance of these effects is glaring given their ubiquity in Nature at all levels: plants, animals and microorganisms base their existence on molecular mechanisms in which the presence of a solvent is essential \cite{jaramillo2011explicit,lovell2002femo, rohrig2005solvent, prabhu2006protein, timasheff1992solvent}.  It is also needless to remark the importance of accounting for solvation effects in almost all branches of Chemistry\cite{chen2009steric, anslyn2006modern,sherwood2019solvent, lucent2007protein}.

Concerning the strategies adopted so far to include solute-solvent interactions we can distinguish two broad classes pertaining to an explicit or implicit treatment of the solvent in the system description. The former is typically represented by Molecular Dynamics simulations or Monte Carlo simulations in which average properties are obtained from sampling the phase space of the system where the solvent degrees of freedom are explicitly accounted for \cite{florova2010explicit, christen2008searching}. This option, due to the high number of molecules needed, is limited to a classical description of the solvent (possibly coupled to a quantum chemical description of the solute in a QM/MM approach\cite{caricato2013vertical, chibani2014modelling}) or to a quantum description for a relatively small systems and limited statistical sampling \cite{cao2005mechanisms, bryantsev2008calculation}. In the long term, one may expect that the development of quantum computers will be sufficiently advanced to allow explicit, full-fledged quantum simulations of solutions. However, to date, this possibility is still far from being realised, and extending implicit solvation models to quantum computing approaches is a suitable option to address solvation effects.

In particular, implicit methods of solvation are the most commonly adopted providing a methodology to describe the surrounding solvent as a continuum medium \cite{roux1999implicit}. Within this framework, the Polarizable Continuum Model (PCM)\cite{tomasi2005quantum} represents {\it de facto} the standard approach due to its flexibility and accuracy. In particular, the Integral Equation Formalism version of the PCM (IEF-PCM) allows to describe with very little modifications to the working equations the presence of both an isotropic or anisotropic polarizable medium as well as ionic solutions \cite{cances1997new, mennucci1997evaluation, cances1998new}. Beyond the environment complexity available, this theoretical framework has been developed in many directions giving the possibility to describe the solute at different levels of theory\cite{cammi2009quantum, cammi2002second, cammi1999linear, caricato2011ccsd}, picturing the overall process with the use of the open quantum system formalism \cite{guido2020open} and also accounting for the presence of nanometallic structures \cite{corni2001enhanced} and optimal control procedures \cite{rosa2019quantum}. Finally, we also mention a recent development which exploits the emerging tool of machine learning to improve the estimates of solvation free-energy obtained from PCM\cite{alibakhshi2021improved}. 

In this contribution we leverage the standard formulation of the IEF-PCM to include solvation effects in the flagship algorithm of quantum simulation for Noisy Intermediate Scale Quantum (NISQ) devices: the Variational Quantum Eigensolver (VQE) \cite{peruzzo2014variational, kandala2017hardware, tilly2022variational}. The choice of this method has been dictated by the recent literature that has showed its successful applications on near-term quantum processors to simulate molecules, condensed matter physics and other phenomena of physico-chemical interest\cite{di2020variational, yoshioka2020variational, pravatto2021quantum, barkoutsos2018quantum, liu2020simulating}. In particular, we exploit a specific flavor of VQE\cite{arrazola2021universal} where the trial wavefunction is built exploiting an adaptive concept\cite{grimsley2019adaptive}.  In the following, we will refer to the new algorithm as PCM-VQE.

This work is organized as follows: first we review the basics of PCMs with particular attention to the IEF-PCM formulation, subsequently we discuss the changes incorporated into the VQE to include solvent effects. In the Results sections we report various numerical tests on three different molecules, namely H$_2$O, BeH$_2$ and H$_3^+$ in dimethyl sulfoxide (DMSO) as a test bed for the algorithm implementation. Further, we assess the estimate of the solvation free energy with a noisy quantum simulation adopting a noise model based on the IBM Q Mumbai quantum processor. We conclude discussing our results and future perspectives of this work.

\section{Theory}

\subsection{The Polarizable Continuum Model}\label{PCM}





The purpose of this section is to recall the PCM concepts and quantities that enter the modifications we have made to the  VQE algorithm. For comprehensive summaries we refer to Refs. \cite{tomasi2005quantum, tomasi1994molecular}.

The physical picture encompassed by PCMs is of a solute embedded in a molecular shaped cavity interacting with the solvent, located outside, which is described as a structureless polarizable dielectric. In this approach, the charge density of the solute molecule polarizes the external environment which generates an electric field (the reaction field) that acts back on the solute. Such reaction field is obtained as the field produced by a set of polarization charges, the so called apparent surface charges (ASC), spread on the cavity boundary, whose values depend, in turn, on the solute molecular electrostatic potential.

To organize the discussion, we first present how such ASC can be calculated within IEF-PCM. Then, we describe how the solute-solvent interaction is accounted for in the quantum mechanical description of the molecule.
\subsubsection{The electrostatic problem}

We start by solving an electrostatic problem in which we look for the electrostatic potential $\varphi(\textbf{r})$ generated by the molecular (nuclear and electronic) charge density $\rho(\textbf{r})$ embedded in a polarizable surrounding solvent characterized by the dielectric constant $\epsilon$. This is accomplished solving the appropriate Poisson's equation \cite{jackson1999classical}:

\begin{equation}\label{poisson_equation}
    \nabla \cdot [\epsilon(\textbf{r}) \nabla \varphi(\textbf{r})] = -4\pi\rho(\textbf{r})
\end{equation}
where $\epsilon(\textbf{r})$ is defined as:

\begin{equation}\label{dielectric_constant_function}
\epsilon(\textbf{r}) =
\begin{cases*}
  1 & inside the cavity \\
  \epsilon        & outside the cavity
\end{cases*}
\end{equation}
Equation \ref{dielectric_constant_function} implies the use of a set of additional boundary conditions to solve the Poisson equation ensuring the continuity of the potential and the electric field at the interface of the cavity \cite{mennucci2008continuum}.

In the framework of IEF-PCM\cite{cances1997new, cances1998new} the electrostatic problem is recasted in an integral equation that directly provides the ASC density:

\begin{equation}\label{ief-pcm}
    \big ( \frac{\epsilon+1}{\epsilon-1} \hat{I} - \frac{1}{2\pi} \hat{D} \big ) \hat{S} \sigma (\textbf{s}) = - \big ( \hat{I} - \frac{1}{2\pi} \hat{D} \big ) \Phi(\textbf{s}) \, . 
\end{equation}
Here $\Phi(\textbf{s})$ is the Molecular Electrostatic Potential (MEP) at the surface $\Gamma$ of the cavity, $\textbf{s}$ is a point on the cavity surface, $\hat{I}$ is the identity operator, $\hat{D}$ and $\hat{S}$ are the components of the Calderòn projector\cite{cances1998new} that are related, respectively, to the normal component w.r.t. $\Gamma$ of the field generated by $\sigma (\textbf{s})$ and the related electrostatic potential at the surface. Their explicit expression depends only on the cavity shape and the dielectric properties of the solvent.\cite{cances1998new}

The numerical solution of Eq. \ref{ief-pcm} involves a discretization of the cavity surface into $N_{tess}$ tesserae and a corresponding discrete representation of the operators $\hat{D}$, $\hat{S}$ and of the ASC density. The formal details of the cavity discretization procedure is described in Refs.\cite{york1999smooth, pascual1990gepol}. Here, we will focus on reporting the working equations of the IEF-PCM method after this step is completed.

The discretization of $\sigma (\textbf{s})$ results in the introduction of a set of charges $\textbf{q}$ positioned at the centre of each tessera:

\begin{equation}\label{sigma_discretized}
    \sigma(\textbf{r}) = \sum_{i=1}^{N_{tess}} \frac{q_i}{a_i} \delta(\textbf{r} - \textbf{s}_i) \, ,
\end{equation}
where $a_i$ and $q_i$ are, respectively, the area and the point charge located at the $i$th tessera, $\delta(\textbf{r} - \textbf{s}_i)$ is a Dirac delta function peaked on the tessera representative point $\textbf{s}_i$. 

Once the discretization procedure has been accomplished we obtain an expression for the polarization charges on all the tesserae, which model the response of the solvent to the presence of the solute:

\begin{equation}\label{discretized_ief_pcm}
    \textbf{q} = - \left( 2\pi \frac{\epsilon+1}{\epsilon-1} {\textbf{S}} - \textbf{D}\textbf{A}\textbf{S} \right)^{-1}  \left( 2\pi  {\mathbf{1}} - \textbf{D}\textbf{A}  \right)\textbf{V}=\mathbf{Q}^{\textrm{PCM}}\textbf{V}
\end{equation}
The equation above is the discretized version of Eq. \ref{ief-pcm}, which gives explicitly $\textbf{q}$ as a function of $\textbf{V}$, which is the vector collecting all the values of the MEP $\Phi(\mathbf{s}_i)$ on the tesserae representative points $\textbf{s}_i$ ($V_i=\Phi(\mathbf{s}_i)$). 

The quantities in bold indicate vectors and matrices that represent the quantities and operators in Eq. \ref{ief-pcm}. Particularly, \textbf{q} and \textbf{V} are column vectors of dimension $N_{\textrm{tess}}$, and \textbf{A} is a diagonal matrix collecting the areas of all the surface elements. $\mathbf{Q}^{\textrm{PCM}}$ is implicitly defined in Eq. \ref{discretized_ief_pcm}, and it is called the solvent response matrix.

So far, we have seen how to obtain both formally (Eq. \ref{ief-pcm}) and practically (Eq. \ref{discretized_ief_pcm}) an expression for calculating the polarization of the solvent (polarization charges, \textbf{q}) due to the presence of the solute (MEP, \textbf{V}).
Let us now see how this impacts the quantum-mechanical description of the molecular system.

\subsubsection{The quantum mechanical problem}

The standard approach for including solvation effects in the quantum description of the molecule is to define a new quantity with respect to which optimize the quantum state of the molecule, such a quantity is known as free energy in solution $\mathcal{G}$:

\begin{equation}\label{free_energy_functional}
    \mathcal{G}[|\Psi\rangle] = \langle \Psi | \hat{H}_{\textrm{0}} + \frac{1}{2} \hat{V}_{\sigma} | \Psi \rangle
\end{equation}
As we can see  $\mathcal{G}$ is a functional of the electronic state only, since we are implicitly adopting the Born-Oppenheimer approximation. In the previous equation, $\hat{H}_{\textrm{0}}$ is the electronic Hamiltonian of the molecule in gas phase and $\hat{V}_{\sigma}$ is the operator accounting for the Coulomb interaction between the ASCs $\sigma({\bf s})$ representing the solvent polarization generated by the molecule's charge density $\rho({\bf s})$ and the electrons of the molecule.


If we apply the variational principle to the free energy functional $\mathcal{G}[|\Psi\rangle]$ (Eq. \ref{free_energy_functional}), under the constraint of a normalized wavefunction, it is possible to derive a non-linear Schr\"odinger equation with an effective Hamiltonian $\hat{H}^{\textrm{eff}}$ that includes the solute-solvent interaction\cite{cammi1995}:

\begin{equation}\label{}
    \hat{H}^{\textrm{eff}}_{|\Psi\rangle} |\Psi\rangle = [\hat{H}_{\textrm{0}} + \hat{V}_{\sigma}(|\Psi\rangle)] |\Psi\rangle = E |\Psi\rangle
    \label{effective_hamiltonian} \, ,
\end{equation}
where we have highlighted the non-linearity of the equation by explicitly reporting the dependence of the interaction operator on the electronic wavefunction.

For the sake of our purposes, is convenient to define the interaction operator in second quantization. This allows us to get a deeper insight into the meaning of this operator and also to illustrate better how to calculate the solvation free energy within the VQE algorithm. 

Therefore, by separating the contributions of electrons and nuclei, in second quantization we can write $\hat{V}_{\sigma}$ as:

\begin{equation}\label{interaction_operator_second_quantization}
    {\color{black}\hat{V}_{\sigma} = W_{NN} + \textbf{v}_N^T \cdot \langle \Psi | \hat{\textbf{Q}} | \Psi \rangle + \sum_{p,q} [\frac{1}{2}(j_{pq} + y_{pq}) + x_{pq}] \hat{E}_{pq} \,.}
\end{equation}
\textcolor{black}{Here $W_{NN}$ is the interaction between nuclei and their polarization charges;} the indices $p$, $q$ run over the basis of molecular orbitals, $j_{pq}$ is the interaction term between the electrostatic potential produced by the electronic charge distribution $-\chi_p(\textbf{r})^*\chi_q(\textbf{r})$, evaluated at each tessera, with the ASC generated by the nuclear charge distribution:

\begin{equation}
    j_{pq} = \textbf{v}_{pq}^T \cdot \textbf{q}_N  \qquad   (\textbf{v}_{pq})_i = -\langle \chi_p | \frac{1}{|\textbf{r} - \textbf{s}_i|} | \chi_q \rangle \, ,
\end{equation}

Similarly, $y_{pq}$ represents the interaction between the nuclear potential and the ASC generated by the elementary electronic charge distribution $-\chi_p(\textbf{r})^*\chi_q(\textbf{r})$, called $\textbf{q}_{pq}$

\begin{equation}
    y_{pq} = \textbf{v}_{N}^T \cdot \textbf{q}_{pq} \qquad (\textbf{v}_{N})_i = \sum_m \frac{Z_m}{|\textbf{R}_m - \textbf{s}_i|} \qquad (\textbf{q}_{pq})_i = \sum_{j} \mathbf{Q}^{\textrm{PCM}}_{ij} (\textbf{v}_{pq})_j \, ,
\end{equation}
where \textbf{v$_N$} is the nuclear potential and $Z_m$, $\textbf{R}_m$ are the nuclear charge and position of the $m$th nucleus.

Finally, we have the interaction term between the electrons and the ASC generated by themselves:

\begin{equation}\label{electron_ASC}
    x_{pq} =  \textbf{v}_{pq}^T \cdot \langle \Psi | \hat{\textbf{Q}} | \Psi \rangle \,,
\end{equation}
where we have introduced the apparent charge operator $\hat{\textbf{Q}}$, also appearing in Eq. \ref{interaction_operator_second_quantization}, given by:

\begin{equation}
    \hat{\textbf{Q}} = \sum_{pq} \textbf{q}_{pq} \hat{E}_{pq} \, .
\end{equation}

The operator $\hat{V}_{\sigma}$ reported in Eq. \ref{interaction_operator_second_quantization} is a one-body operator since it represents the interaction between a charge distribution (the ASC) and the electrons  of the molecule, formally analogous to the interaction term between nuclei and electrons in the standard molecular Hamiltonian. Since it is a spin-free operator, we have written it directly in terms of singlet excitation operators $\hat{E}_{pq}$:

\begin{equation}
    \hat{E}_{pq} = a^{\dagger}_{p\alpha} a_{q\alpha} + a^{\dagger}_{p\beta} a_{q\beta}
\end{equation}

To conclude this section we summarize the standard procedure to find the solution of the coupled equations for the solvent (Eq. \ref{discretized_ief_pcm}) and the solute (Eq. \ref{effective_hamiltonian}) response.

The idea is to find the minimum of the $\mathcal{G}$ functional with a self-consistent procedure. For a given initial approximation of the many-electron wave function of the molecule, the electrostatic potential \textbf{V} is calculated on each tessera. Subsequently, the polarization charges are obtained using (Eq. \ref{discretized_ief_pcm}). In turn, such charges enter directly the definition of the effective Hamiltonian (see Eq. \ref{interaction_operator_second_quantization}) that allows us to compute the an improved wavefunction and the corresponding $\mathcal{G}$, (Eq. \ref{free_energy_functional}). Then, one iterates these steps to converge the value of the free energy.

In the next section we will describe the PCM-VQE algorithm. At the heart of this new hybrid quantum-classical algorithm there is the idea of translating the just mentioned self-consistent procedure to a procedure where the minimization of the free energy functional and the solution of the electrostatic problem are performed classically while the quantum computer is used to generate the trial wavefunction and evaluate the expectation values needed to calculate the corresponding solvation free energy.

\section{PCM-VQE}

Here we describe the extension of the VQE algorithm to include the solvation effects using the PCM model described in the previous section.

We start considering the standard workflow of the VQE. We use a quantum computer to prepare a trial state of the N-electrons molecular wave function and to measure the expectation value of the corresponding Hamiltonian. Subsequently, the prepared state is variationally optimized to find the ground state energy. A classical optimizer is used to adjust the variational parameters $\bar{\theta}$ that define the quantum circuit preparing the many-electron wave function.

In order to account for solvation effects within the VQE algorithm, we generalize the objective function to be the free energy in solution as defined in Eq. \ref{free_energy_functional}: 

\begin{equation}\label{parametrized_free_energy}
    \mathcal{G}[\bar{\theta}] = \langle \hat{H}_{\textrm{0}} \rangle_{\bar{\theta}} + \frac{1}{2} \langle \hat{V}_{\sigma} (\bar{\theta}) \rangle_{\bar{\theta}}
\end{equation}
Where we recall $\hat{H}_{\textrm{0}}$ is the molecular Hamiltonian of the molecule \textit{in vacuo} and $V_{\sigma}$ is the solute-solvent interaction operator defined in Eq. \ref{interaction_operator_second_quantization}. By taking the expectation values of these observables in the prepared state $\vert \Psi(\bar{\theta})\rangle$ we re-write the cost function as,

\begin{equation}\label{eq:gpcm}
    \mathcal{G}[\bar{\theta}] = \sum_{p,q} h_{pq} d_{pq}(\bar{\theta}) + \frac{1}{2} \sum_{p,q,r,s} g_{pqrs} D_{pqrs}(\bar{\theta}) + \frac{1}{2} \sum_{p,q} [(j_{pq} + y_{pq}) + x_{pq}(\bar{\theta})] d_{pq} (\bar{\theta}) + \frac{1}{2}W_{NN}
\end{equation}
Here $p,q,r,s$ are indices running over the orbitals (note we are considering a spin free Hamiltonian, proper for the usual condition when no magnetic fields or spin-orbit coupling is considered), $d_{pq}(\bar{\theta})$ and $D_{pqrs}(\bar{\theta})$ are the one- and two-electrons orbital Reduced Density Matrices (1-, 2-RDMs) defined as follows\cite{helgaker2014molecular}:

\begin{equation}\label{rdms}
\begin{split}
        d_{pq}(\bar{\theta})& = \langle E_{pq} \rangle_{\bar{\theta}} \\
        D_{pqrs}(\bar{\theta})& =  \langle E_{pq}E_{rs} - \delta_{rq}E_{ps} \rangle_{\bar{\theta}} \, .
\end{split}
\end{equation}
The possibility of retrieving  $d_{pq}(\bar{\theta})$ and $D_{pqrs}(\bar{\theta})$ as expectation values is guaranteed by the inherent variational procedure of the method, which is also the case for the UCCSD ansatz \cite{taube2006new}. This is non-trivial in general, and for non variational approaches the expression should be replaced by strategies such as the introduction of an auxiliary variational Lagrangian. In those cases, the use of Eq. \ref{rdms} would represent just an approximation\cite{helgaker2014molecular} to the proper density matrices.

From what we have seen in the previous section it is easy to see that the solvation free energy contribution to the total free energy depends both implicitly and explicitly on the circuit parameters. The implicit dependency stems from the definition of the new interaction operator, the explicit dependency is a result of the relaxation of the wavefunction in presence of the reacting field. 
Concerning the dependence of the interaction operator matrix elements on the variational parameters, it is instructive to get a better intuition on the modified procedure of the PCM-VQE to make explicit the presence of the variational parameters in Eq. \ref{electron_ASC}:

\begin{equation}
    x_{pq} =  \textbf{v}_{pq}^T \cdot \langle \Psi | \hat{\textbf{Q}} | \Psi \rangle \xrightarrow{\textrm{PCM-VQE}} x_{pq}(\bar{\theta}) =  \textbf{v}_{pq}^T \cdot \langle \Psi (\bar{\theta}) | \hat{\textbf{Q}} | \Psi (\bar{\theta}) \rangle
\end{equation}
This last feature gives rise to an hybrid algorithm in which the classical optimization routine is tasked with the optimization of a cost functional with a dependence on the parameters that is different to that of standard VQEs: as a consequence of the non-linearity here we jointly optimize the quantum state and the observable w.r.t. which we compute the expectation value.
 Whether this feature has an impact on the convergence properties of the algorithm is a topic that deserves further study in terms of the theory of hybrid variational algorithms per se. In this paper, we will address this problem numerically only for a few examples.

Another important point to comment is that with the definition of this new cost functional we are including solvent effects in our description without any additional cost of quantum computational resources \textcolor{black}{(see the SI Sec. "Algorithmic complexity: PCM overhead" for a more in-depth analysis of the computational cost)} as the same quantities needed to measure the Hamiltonian expectation value in gas-phase are needed to update the interaction operator matrix elements (as shown in Eq. \ref{discretized_ief_pcm} and Eq. \ref{electron_ASC} they only depend on the 1-RDM) and to compute the solute-solvent contribution to the free energy (Eq. \ref{parametrized_free_energy}).

Finally, we notice that once the density matrices are extracted from the QC, the solution of the electrostatic problem using Eq. \ref{discretized_ief_pcm} is straightforward as the solvent response matrix $\textbf{Q}^{\textrm{PCM}}$ remains unchanged through all the calculation.

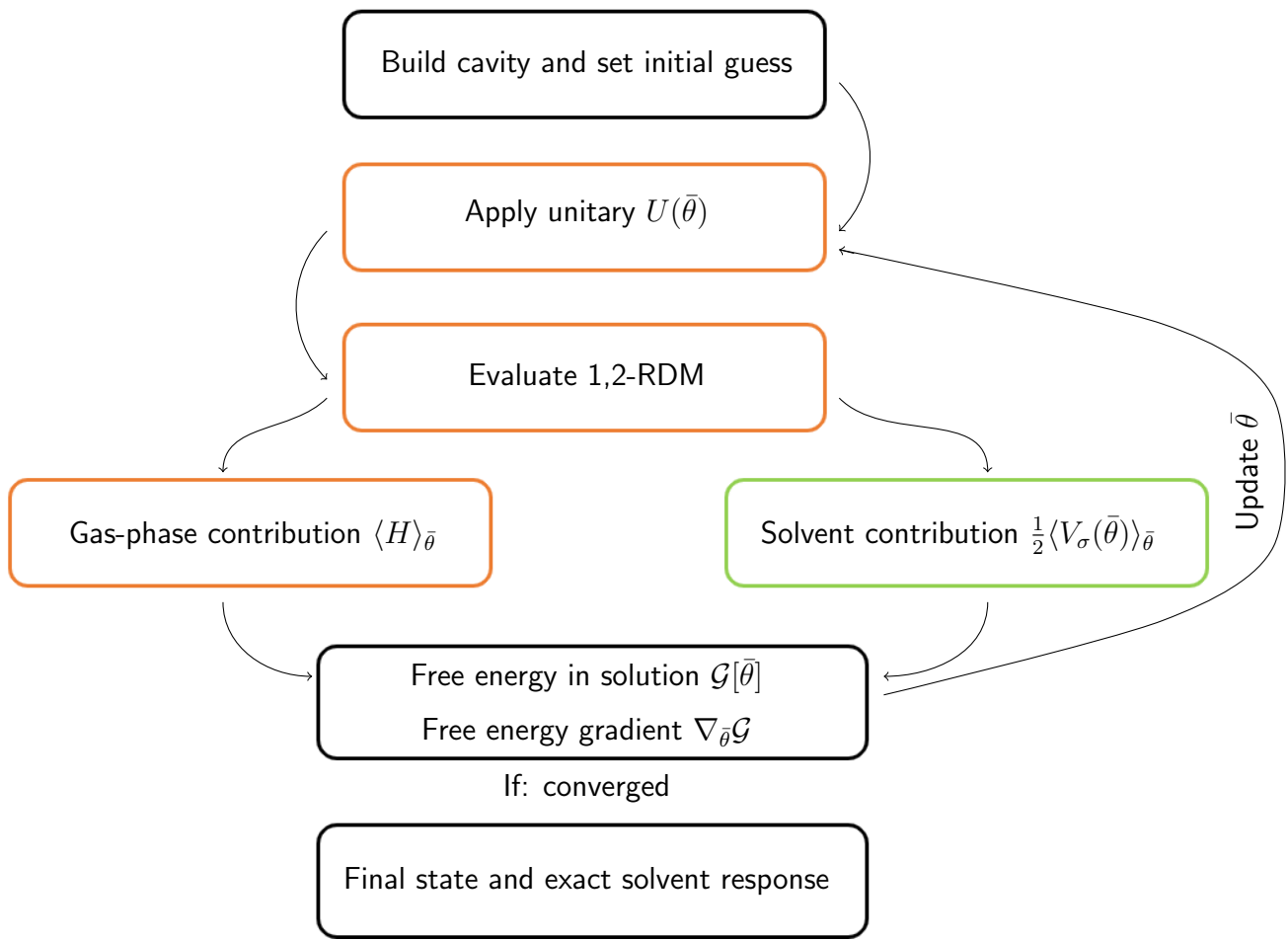
\begin{figure}[h!]

\begin{tikzpicture}[node distance=cm,
    every node/.style={fill=white, font=\sffamily}]

    \node (figure) at (0,0) {\centering
    \includegraphics[width = \textwidth]{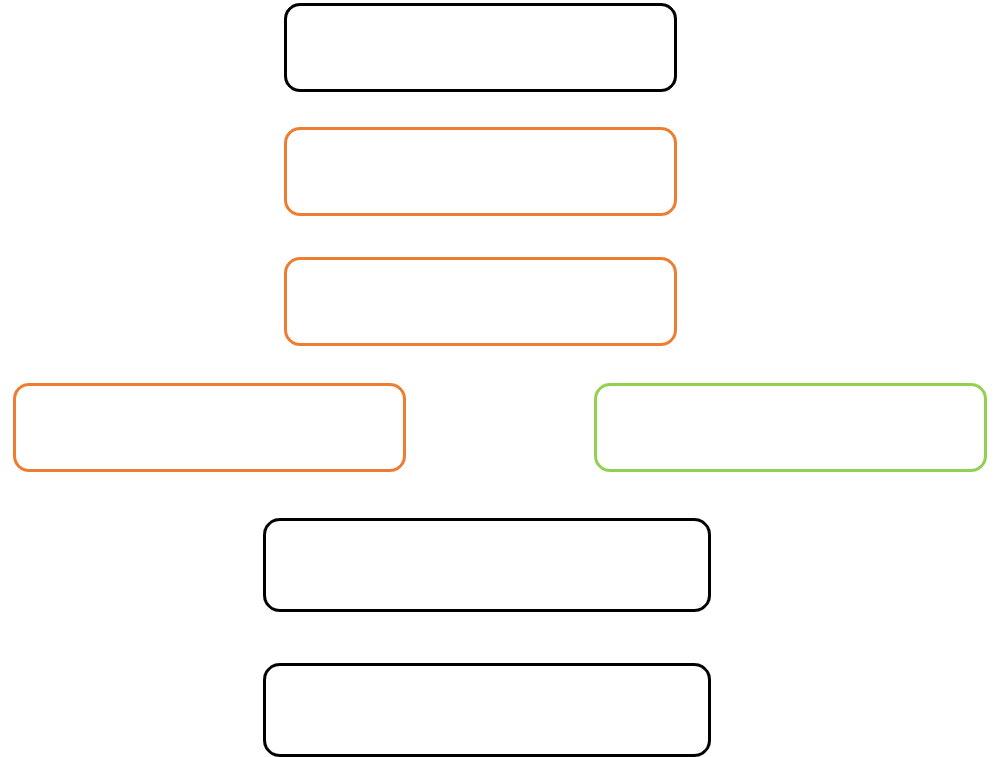}};
    
    \draw [->] {(3.15, 5.25) to [out=-45, in = 45] (3.15, 3.25) };
    
    \draw [->] {(-3.75, 3.25) to [out=-135, in = 135] (-3.75, 1.25) };
    
    \draw [->] {(-3.75, 1) to [out=-135, in = 90] (-5.15, 0) };
    
    \draw [->] {(3.15, 1) to [out=-45, in = 90] (5.15, 0) };
    
    \draw [->] {(-5.15, -1.75) to [out=-90, in = 180] (-3.95, -2.75) };
    
    \draw [->] {(5.15, -1.75) to [out=-90, in = 0] (3.75, -2.75) };
    
    \draw plot [smooth] coordinates {(3.75, -3) (7.5, -2) (9, -1) (9, 1) (7.5, 2) (3.15, 3)};
    
    \draw [->] {(3.35, 2.95) to (3.15, 3)};
    
    \node (a) at (-0.25, 5.5) {Build cavity and set initial guess};
    
    \node (b) at (-0.25, 3.5) {Apply unitary $U(\bar{\theta})$};
    
    \node (c) at (-0.25, 1.30) {Evaluate 1,2-RDM};
    
    \node (d) at (-4.75, -0.85) {Gas-phase contribution $\langle H \rangle_{\bar{\theta}}$}; 
    
    \node (e) at (4.75, -0.85) {Solvent contribution  $\frac{1}{2} \langle V_{\sigma} (\bar{\theta}) \rangle_{\bar{\theta}}$}; 
    
    \node (f) at (-0.25, -2.75) {Free energy in solution $\mathcal{G}[\bar{\theta}]$};
    
    \node (l) at (-0.25, -3.5) {Free energy gradient $\nabla_{\bar{\theta}}\mathcal{G}$};
    
    \node (g) at (-0.25, -5.5) {Final state and exact solvent response};

    \node (h) at (-0.25, -4.25) {If: converged};    
    
    \node (i) at (8.65, 0) [rotate = 90] {Update $\bar{\theta}$};

\end{tikzpicture}
    \caption{Schematic representation of the PCM-VQE algorithm. Black boxes represent operations that involve uniquely classical computation, orange boxes refer to operations that are performed by the quantum computer. Computing the solvent response (green box) is an hybrid computing operation as the classical solver of the IEF-PCM equation is fed by the quantum processor with the 1-RDM. The PCM-VQE loop is iterated until a  convergence criterion is satisfied providing the final state of the solute molecule and the corresponding reaction field of the solvent. }
    \label{PCM-VQE_scheme}
\end{figure}

The PCM-VQE algorithm consists of five steps, as it is sketched in Fig. \ref{PCM-VQE_scheme}:

\begin{enumerate}
    \item The molecular wavefunction is encoded into the state of the quantum computer. Several mappings have been developed in the literature, see Ref.\cite{cao2019quantum} for an extended review on the topic.
    \item Transform the quantum computer initial state according to a unitary operation (often referred as the VQE ansatz) $U(\bar{\theta})$ which depends parametrically on the set parameters $\bar{\theta}$.
    \item  Evaluate the one- and two-electron orbital RDMs using the trial state prepared by the quantum computer.
    \item Update the polarization charges using Eq. \ref{discretized_ief_pcm} and the matrix elements $j_{pq}$, $y_{pq}$ and $x_{pq}(\bar{\theta})$ accordingly.
    \item Evaluate the free energy functional $\mathcal{G}[\bar{\theta}]$ and compute, if needed, its gradient with respect to the circuit parameters.
\end{enumerate}

Finally, steps 2-5 are repeated until the value of the molecule's free energy in solution is converged.

For the sake of clarity, here we stress that the non-linearity of the effective Hamiltonian in Eq. \ref{effective_hamiltonian} prevents the inclusion of the solvation effects in the VQE by simply substituting the molecular integrals computed in gas-phase with the molecular integrals computed in solution.

Indeed, the result obtained in this fashion would only provide the optimal variational parameters which enable to prepare the ground state of a molecule \textit{in vacuo} whose MOs result from a HF calculation in solution. Such a wavefunction would differ (providing inaccurate results) both from the solution given by the IEF-PCM model coupled to a standard method of quantum chemistry and from the PCM-VQE. A scheme of the algorithm highlighting the interplay of classical and quantum libraries is given as Fig. S\ref{PCM-VQE_scheme}; the code is available on GitHub\cite{pcm_vqe_code}.

This concludes the description of the PCM-VQE algorithm and the Theory section. In the following we will discuss the technical details of the implementation and the results obtained both with a noiseless simulation and in presence of a simulated quantum noise.

\subsection{Computational details}

The PCM-VQE algorithm has been implemented in a Python code\cite{pcm_vqe_code} realizing the interface between the Psi4\cite{smith2020psi4} quantum chemistry package and the PennyLane quantum library\cite{bergholm2018pennylane}. Psi4 was used to compute the molecular integrals, build the solute cavity and solve the electrostatic problem (through its interface with PCMSolver\cite{di2019pcmsolver}), and PennyLane functionalities were used to implement the quantum algorithm. That is, defining the quantum circuit preparing the molecular trial state, computing the expectation value of the many-body observables and optimizing the quantum circuit parameters.

\begin{figure}[h!]
\begin{subfigure}{.5\textwidth}
  \centering
  \caption{}
  \includegraphics[width=\linewidth]{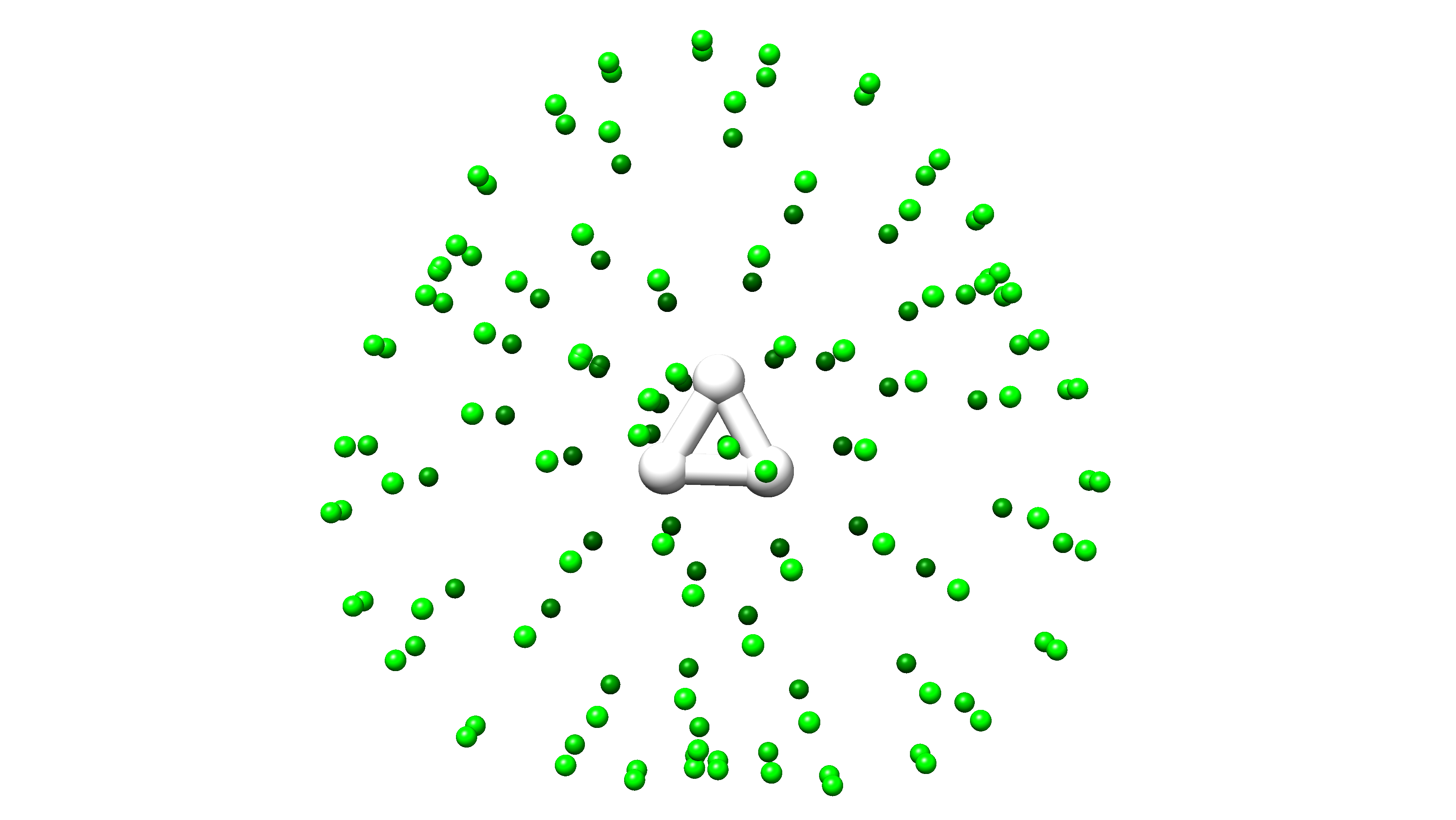}
 
  \label{fig:sfig1}
\end{subfigure}%
\begin{subfigure}{.5\textwidth}
  \centering
  \caption{}
  \includegraphics[width=\linewidth]{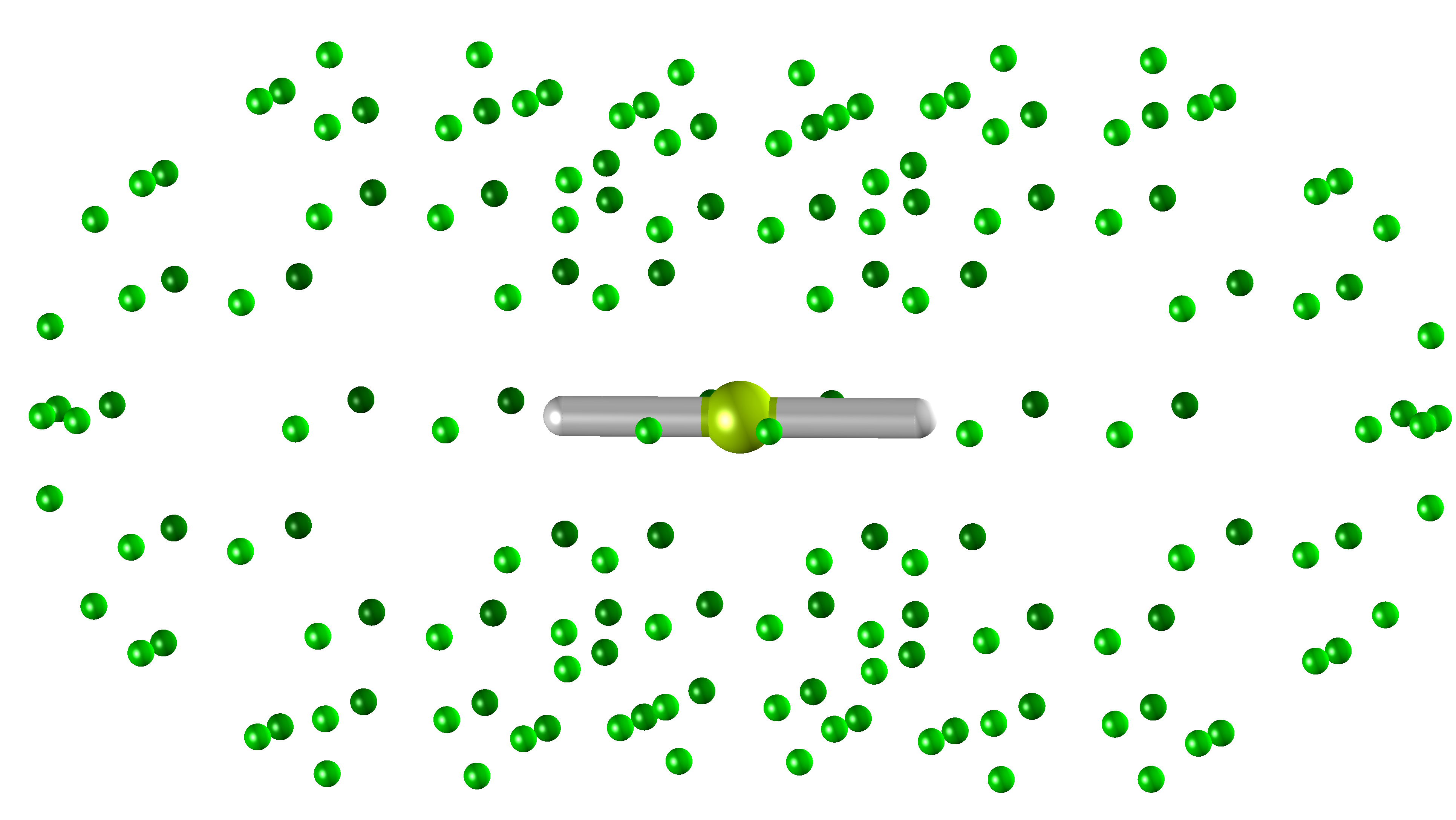}
 
  \label{fig:sfig2}
\end{subfigure}
\begin{subfigure}{.5\textwidth}
  \centering
  \caption{}
  \includegraphics[width=\linewidth]{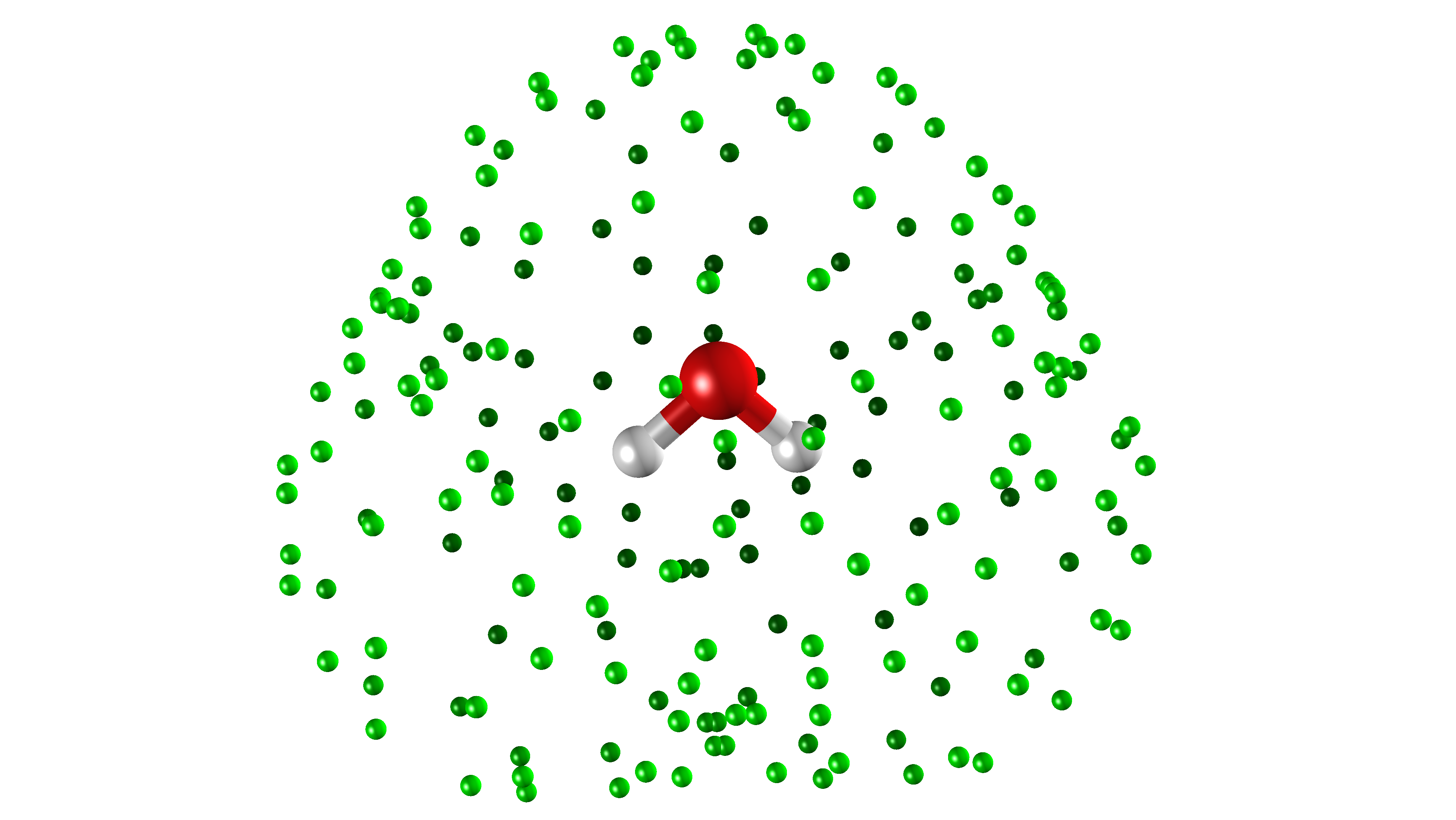}
  
  \label{fig:sfig1}
\end{subfigure}%
\caption{Structures of studied systems: (a) trihydrogen cation, (b) beryllium hydride and (c) water. The small green dots are the representative points of each tessera, where ASCs are located, and are spread on the molecularly shaped cavity boundary. Please note that different cavity sizes are not to scale. These figures have been produced with the software Chimera\cite{pettersen2004ucsf}.}
\label{molecules}
\end{figure}

We have performed numerical simulations to compute the free energy in solution of the trihydrogen cation (H$_3^+$), beryllium hydride (BeH$_2$) and water (H$_2$O) molecules at their equilibrium geometry, \textcolor{black}{shown in Fig.~\ref{molecules}}, computed in gas-phase using the STO-3G basis set. \textcolor{black}{Two examples using a larger basis set (6-31G) are provided in the Supporting Information. Here we remark that the use of STO-3G as basis set should be avoided when the goal of the numerical simulation is to quantitavely predict a property and/or compare the result with an experimental measure. This was not the case for the present study where the purpose of the numerical experiments is to showcase the newly developed algorithm on a set of different molecules. The choice of a minimum basis set is therefore motivated to avoid overflowing the computational resources required by the used quantum computer simulators as previously done in other works\cite{kandala2017hardware, nam2020ground, ratini2022wave}}. The molecular cavities are built in PCMSolver according to the GePol algorithm \cite{pascual1990gepol} using the atomic radii reported in Ref. \cite{bondi1964van}. The choice of the investigated systems has been made to span a set of molecules with different dipole moment, charge state (quantities that are deeply involved when solvation effects are taken into account) and spatial symmetry, so as to test the implementation and algorithmic robustness over different situations. \textcolor{black}{The classical reference calculations have been performed with the Psi4 code at the CCSD/IEF-PCM level of theory\cite{cammi2009quantum} using the same solute cavities.}

\begin{figure}[h!]

\includegraphics[scale = 0.5]{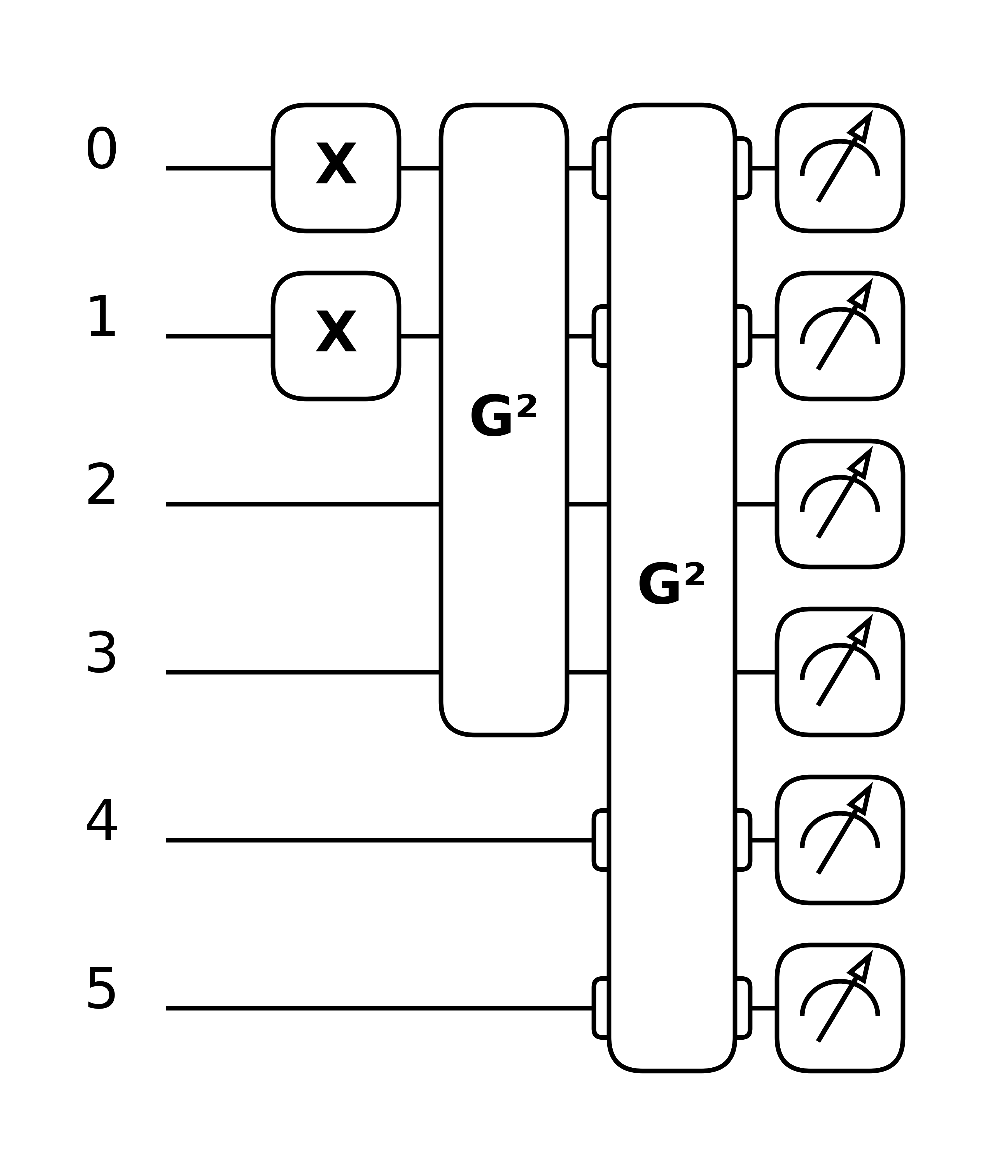}
\caption{Quantum circuit used for the PCM-VQE simulations on the H$_3^{+}$ molecule. As mentioned in the main text, the circuits used are composed by a set of initial state preparation gates (here the X gates acting on the first two qubits) and by particle-conserving unitary operators here implemented as Givens rotations (whence the label G). Figure obtained using the quantum circuit drawer function as implemented in PennyLane\cite{bergholm2018pennylane}.}
\label{circuit_h3+}
\end{figure}

The variational quantum circuits to prepare the trial states $|\Psi(\bar{\theta})\rangle$ of the simulated molecules are built following the methodology reported by the authors in Ref.~\cite{delgado2021variational}. For the sake of completeness, we outline here the main steps for building the quantum circuit. First, the $N$-qubit register encoding the molecular spin-orbitals is initialized to the Hartree-Fock (HF) state of the molecule. That is, the first $N_e$ qubits, with $N_e$ being the number of electrons, are set in the state $\vert 1 \rangle$ and the other $N-N_e$ qubits in the state $\vert 0 \rangle$. Thus, to prepare a many-electron state beyond the meanfield approximation, we apply particle-conserving single- and double-excitation gates on the initial state (see Fig. \ref{circuit_h3+} and SI for the circuits used in this work). These excitation operations are implemented in PennyLane as Givens rotations which act on the subspace of two and four qubits~\cite{arrazola2021universal}. As a result, the final state is a superposition of the HF state and other states encoding multiply-excited configurations~\cite{delgado2021variational}. In \, Tab.\ref{numerical_details} we report the number of variational parameters (i.e., number of Givens rotations) for the systems studied in this work and the corresponding number of maximum iterations needed to achieve the results shown in Fig.\ref{adaptive_comparison} .

\begin{table*}
    \small
    \centering
    \caption{Variational parameters and maximum number of iterations for the PCM-VQE calculations showed in Fig.\ref{adaptive_comparison}.}
    \label{tab:table_1}
    \begin{tabular*}{\textwidth}{@{\extracolsep{\fill}}lccc}
        \hline
        \hline
        & H$_3^+$ & BeH$_2$ & H$_2$O \\
        Variational parameters & 2 & 18 & 30 \\ 
        Max. iterations &  12 & 15 & 12 \\ 
    \end{tabular*}
    \label{numerical_details}
\end{table*}

Furthermore, we have used the adaptive method proposed in Ref.~\cite{delgado2021variational} to select the excitation operations that are important to compute the ground state of the solvated molecules. In addition, to check the reliability of the method to different implementations of VQE, we also explored a more system-agnostic ansatz,  the unitary coupled-cluster ansatz truncated at the level of single and double excitations (UCCSD) \cite{peruzzo2014variational} (see SI).

On the other hand, evaluating the cost function defined in Eq.~\eqref{eq:gpcm} for given set of the variational parameters $\bar{\theta}$ requires to compute the one- and two-electron orbital reduce density matrices $d_{pq}(\bar{\theta})$ and $D_{pqrs}(\bar{\theta})$, respectively. To that aim, we use the Jordan-Wigner transformation~\cite{seeley2012bravyi} to map the fermionic operator $\hat{E}_{pq}$ into the Pauli basis, and compute their expectation values in the trial state prepared by the quantum circuit. Thus, we proceed to minimize the cost function to obtain the free energy of the solvated molecules. The optimization of the circuit parameters in the absence of noise was performed using an adaptive gradient descent algorithm while a gradient-free optimizer~\cite{ostaszewski2021structure} was used in the case of noisy simulations.


In addition, we investigated the capability of the present PCM-VQE implementation to estimate solvent effects in a system in the presence of a high degree of static correlation such as the double dissociation bonding profile of water. For the latter calculations we have used the same variational ansatz exploited for the single point calculations at the equilibrium geometry.

The results concerning the implementation on a simulated noisy quantum hardware are obtained by using a noise model for the IBM Q Mumbai quantum processor as implemented in the Qiskit library\cite{Qiskit}. It includes one- and two-qubits gate error probabilities, pulse duration for the basis gates, readout errors and thermal relaxation effects tuned upon the experimental parameters. Each circuit has been repeated 8192 times to build relevant statistics, we set the number of shots per circuit to match the maximum number allowed on the actual quantum device. \textcolor{black}{In the SI (Sec. "Measurement budget allocation and statistical errors") we provide a theoretical discussion on the error due to a finite sampling of the expectation value. Here we summarize from a practical point of view. The error bars reported in the results' section have been obtained assuming the standard deviation on the expectation value of each Pauli string equal to 1, which is an upper bound for this quantity. Subsequently the error on each Pauli string is obtained dividing by the square root of the number of shots executed (8192 in our case), since their expectation value is obtained by independent measures. The final error bars on (free) energies are then given by standard error propagation applied to the Hamiltonian mapped on the Pauli strings.}

\section{Results}
\label{results}

\subsection{Numerical simulations in noise free conditions}

\subsubsection{Single point calculations}

\begin{figure}[h!]

\begin{tikzpicture}[node distance=cm,
    every node/.style={fill=white, font=\sffamily}]
    
    \node (figure) at (0,0) 
    {\centering
    \includegraphics[width = \textwidth]{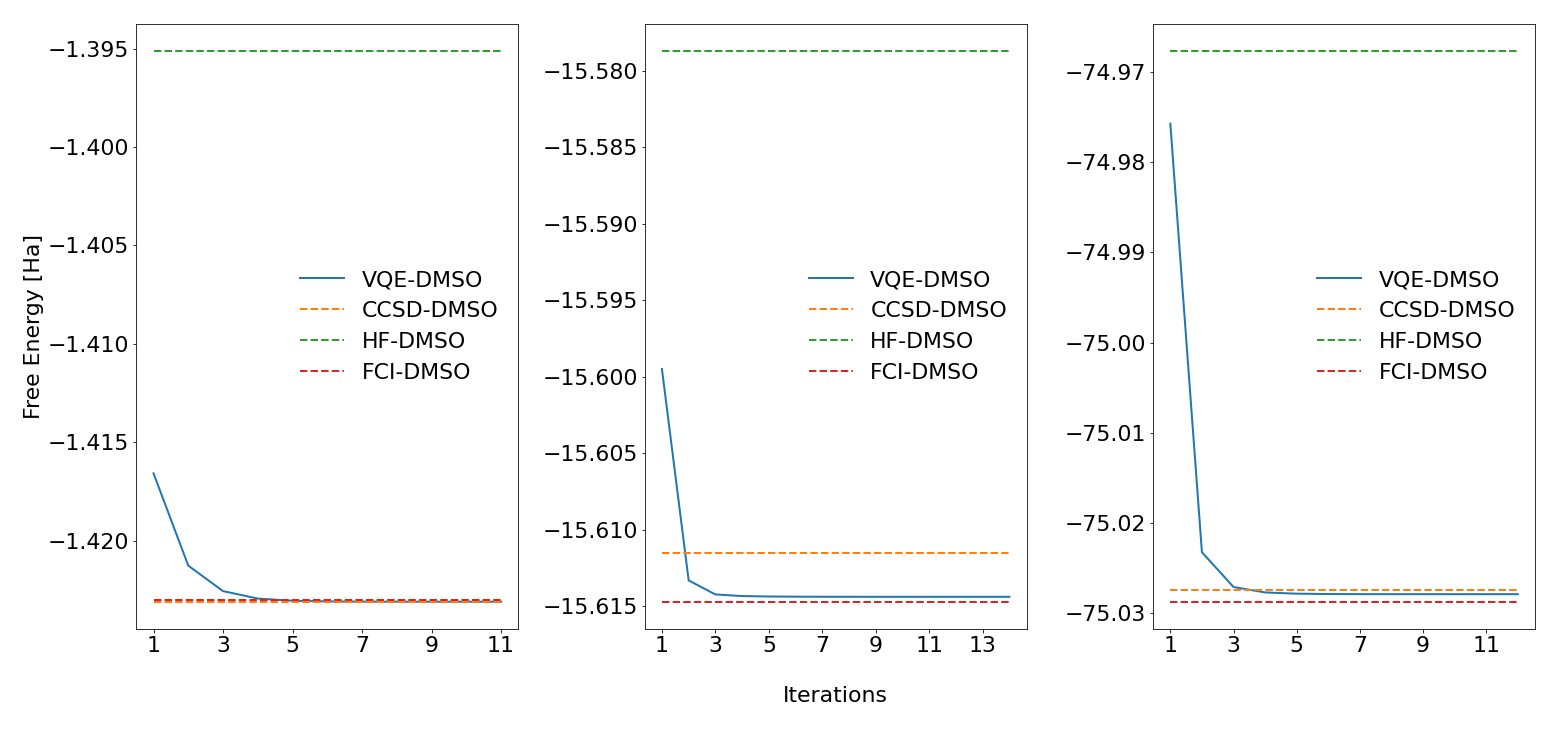}};

\node (a) at (-6.1, 1.75) {(a)};

\node (b) at (-0.55, 1.75) {(b)};

\node (c) at (4.95, 1.75) {(c)};

\end{tikzpicture}
    \caption{PCM-VQE results. Ground state free energies in dimethyl sulfoxide (DMSO) for the (a) Trihydrogen ion H$_3^+$, (b) Beryllium hydride BeH$_2$ and (c) Water H$_2$O. The solid blue line represents the free energy in solution obtained with the PCM-VQE as a function of the iterations using a STO-3G basis set. As a reference we show the free energy obtained with an HF-PCM calculation (green dashed line) and with a CCSD-PCM calculation (orange dashed line) using the same basis set. The red dashed line shows the reference value obtained running a FCI calculation \textit{in vacuo} and adding to the total energy contribution the polarization energy.}
\label{adaptive_comparison}
\end{figure}
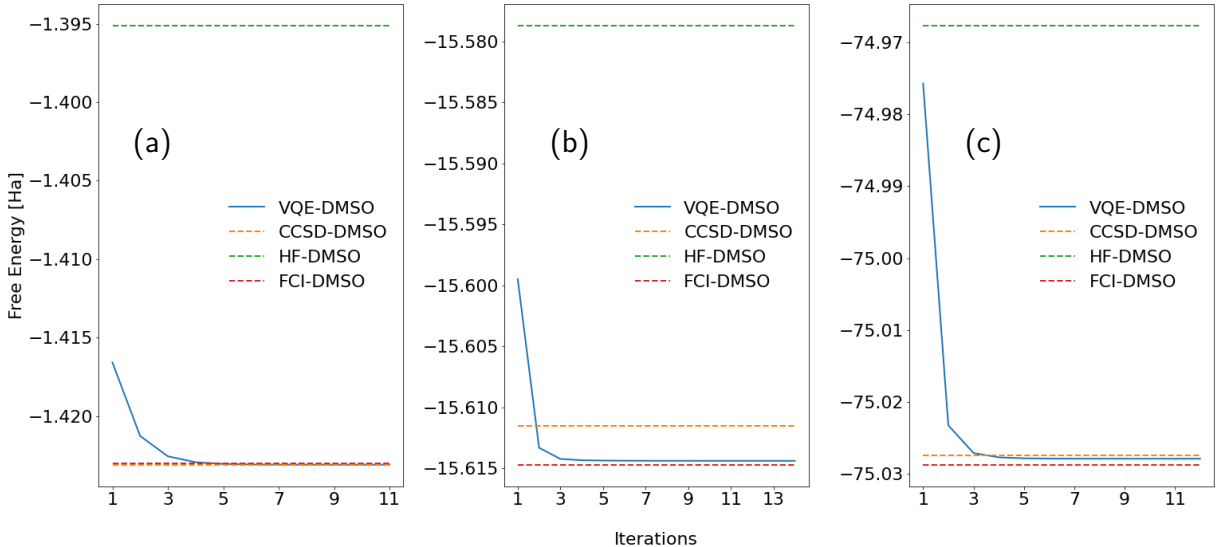

Fig. \ref{adaptive_comparison} shows the values of the free energy in solution $\mathcal{G}[\bar{\theta}]$ as a function of the iterations for the molecules depicted in Fig. \ref{molecules}. All the results of this section are reported following the same color code in the plots. The solid blue line corresponds to the free energy evaluated with the PCM-VQE algorithm, the orange dashed line refers to the free energy evaluated classically at the PCM-CCSD level of theory, and the green dashed line gives the free energy value computed classically with the PCM-HF method. The left panel (Fig. \ref{adaptive_comparison}a) refers to the trihydrogen cation; a two electron system whose wavefunction in the STO-3G basis can be encoded by using six qubits.

This circuit is able to prepare the parameterized state $|\Psi(\theta_1, \theta_2)\rangle$ defined as:
\begin{equation}
    |\Psi(\theta_1, \theta_2)\rangle = \text{cos}(\theta_1)\text{cos}(\theta_2)|\text{HF}\rangle - \text{cos}(\theta_1)\text{sin}(\theta_2)a^{\dagger}_5a^{\dagger}_4a_1a_0|\text{HF}\rangle - \text{sin}(\theta_1)a^{\dagger}_3a^{\dagger}_2a_1a_0|\text{HF}\rangle \, . 
\end{equation}
  Particularly, the last expression shows the efficiency of the adaptive procedure which enables to generate a variational ansatz that spans selectively the subspace corresponding to the set of Slater determinants that contribute to the FCI wavefunction without allowing to reach states having components along different electronic configurations.

As a first comment, we can notice how the quantum simulation algorithm is able to recover all the correlation energy and solvent effects contribution w.r.t. the value given by the CCSD reference which is exact in this case. Moreover, we can see that the optimization convergence is reached within only ten iterations. This is due to the optimization settings that comprise an educated guess encoding the $|\mathrm{HF}\rangle$ state. Furthermore, we have adopted a variational ansatz that prepares the FCI ground-state for this molecule. Such a strategy allows to further reduce the cost of including solvation effects: in the PCM-VQE theory section we have seen that no other additional costs are present concerning the quantum part of the algorithm. Here we point out that only a few more iterations are needed to account for effects of the solvent. In this regard, we highlight that constructing the variational ansatz with an adaptive procedure on the wavefunction in gas-phase is an approximation that applies best if the electronic structure of the solute is not severely modified by the inclusion of the solvent. When this occurs, it may happen that contributions from a few excited configurations, not relevant in the electronic structure \textit{in vacuo}, are lost. In such cases, the resulting wavefunction remains a good guess for the effective Hamiltonian in solution, and the adaptive procedure may be restarted with such Hamiltonian to include the relevant excitations. \textcolor{black}{In the SI we show additional results for a different system (HeH$^+$) in which a more system agnostic ansatz is considered (UCCSD) with both STO-3G and 6-31G basis sets to further assess the performance of the modified algorithm. Moreover, we include an additional calculation with an adaptive circuit for the H$_3^{+}$ molecule using the 6-31G basis set (SI Sec. "H$_3^+$ calculations with PCM-VQE/6-31G")}.

Figs. \ref{adaptive_comparison}b-\ref{adaptive_comparison}c plot the numerical results for the BeH$_2$ and the H$_2$O molecules. For these larger systems the core electrons localized in the s-type orbitals of the beryllium and the oxygen atoms are excluded from the active space. That means that we have four and eight active electrons in the BeH$_2$ and the H$_2$O molecules, respectively, whose wavefunctions are represented using twelve qubits. We chose to keep the core electrons frozen for practical purposes as their optimization does not impact effectively neither the electronic structure nor the description of the solute-solvent interaction.

As we can see looking at the figures all the consideration made above in the case of H$_3^+$ still hold. This implies that, at least for the simple systems that we can tackle with current NISQ devices, the procedure involving first the simulation in absence of the solvent to detect the more relevant excited configurations retains its effectiveness as the size of the system increases. In addition, we can note that in this case, where more electrons are involved, the PCM-VQE algorithm predicts lower energy states as compared with the classical simulations at the level IEF-CCSD. This is not surprising as the variational ansatz spans a larger space than CCSD. Further, this is in accordance with the comparison, \textit{in vacuo}, between the FCI and CCSD wavefunctions, Tab. \ref{free_energy_comparison_data_dmso}.

\begin{table*}
    \small
    \centering
    \caption{Free energies in solution for the studied systems (Fig. \ref{molecules}). Comparison between FCI/CCSD/VQE (gas-phase) and between IEF-FCI/IEF-CCSD/PCM-VQE (DMSO). Molecular geometries for the calculations as well as values obtained for the VQE in gas phase with the adaptive ansatz have been taken from Ref. \cite{delgado2021variational}. Energy values are reported in Hartree.}
    \label{tab:table_1}
    \begin{tabular*}{\textwidth}{@{\extracolsep{\fill}}lccc|ccc}
        \hline
        \hline
    &\multicolumn{3}{c}{Gas-phase}&\multicolumn{3}{c}{Solvent}\\
    \cline{2-7}
        & FCI & CCSD & VQE & FCI-PCM$^{a}$ & CCSD-PCM & PCM-VQE \\
        $\mathcal{G}_{\text{H}_3^+}$ & -1.2744 & -1.2744 & -1.2744 & -1.4230 & -1.4231 & -1.4231 \\ 
        $\mathcal{G}_{\text{BeH}_2}$ &-15.5952 & -15.5945 & -15.5945 & -15.6147 & -15.6115 & -15.6144  \\ 
        $\mathcal{G}_{\text{H}_2\text{O}}$ & -75.0233 & -75.0231 & -75.0230 & -75.0287 & -75.0273 & -75.0279 \\ 

    \end{tabular*}
    \subcaption{The results reported under the label FCI-PCM are obtained by adding to the FCI energy in gas phase the polarization energy computed with the 1-RDM of the corresponding wavefunction.}
    \label{free_energy_comparison_data_dmso}
\end{table*}

We observe from Tab. \ref{free_energy_comparison_data_dmso} that the variational ansatz is able to recover, in gas-phase, almost all the correlation energy in all cases with maximum differences of $\approx$ 0.4 kcal/mol (i.e., within chemical accuracy). Further, moving to the calculations in DMSO, with the PCM-VQE we are able to improve the results of the IEF-CCSD wavefunctions up to 1.9 kcal/mol as in the case of BeH$_2$. For the sake of completeness we reported in Tab. \ref{solvation_data} the solvation free energies for the simulated molecules in DMSO.
We recall that the solvation free energy is computed by taking the difference of the free energy in solution and the energy of the solute \textit{in vacuo}:

\begin{equation}
    \Delta \mathcal{G}_{sol}= \mathcal{G}[\bar{\theta}_{sol}] - \langle \Psi(\bar{\theta}_{vac}) | H | \Psi(\bar{\theta}_{vac}) \rangle
    \label{eq:sol}
\end{equation}
 While in the next section we will look at this quantity evaluated in presence of quantum noise, here we focus on the accuracy obtained with the noiseless calculations. In all cases we are able to recover quantitatively the description obtained with a IEF-CCSD calculation with the larger deviation observed for the BeH$_2$ molecule. 

\begin{table*}
    \small
    \centering
    \caption{Solvation free energies for the studied systems (Fig. \ref{molecules}). Comparison between IEF-CCSD and PCM-VQE in DMSO. Energy values are reported \textcolor{black}{in eV}.}
    \label{tab:table_1}
    \begin{tabular*}{\textwidth}{@{\extracolsep{\fill}}lccc}
        \hline
        \hline
        & H$_3^+$ & BeH$_2$ & H$_2$O \\
        $\Delta\mathcal{G}_{\text{PCM-VQE}}$ & \textcolor{black}{-4.046} & \textcolor{black}{-0.539} & \textcolor{black}{-0.133} \\ 
        $\Delta\mathcal{G}_{\text{IEF-CCSD}}$ &  \textcolor{black}{-4.046} & \textcolor{black}{-0.460} & \textcolor{black}{-0.114} \\ 
    \end{tabular*}
    \label{solvation_data}
\end{table*}

In the SI (Tab. S1) we have also reported analogous results obtained for the same molecules in water.

\subsubsection{Solvation effects along the water molecule double bond dissociation profile}

In this section we apply the PCM-VQE to estimate solvation effects along the water molecule double bond dissociation profile. Such a system has been largely investigated in the context of electronic structure theory \cite{yanai2006canonical}\cite{van2000benchmark}\cite{evangelista2011orbital} and very recently it has been considered in Ref.\cite{mizukami2020orbital} to benchmark the effects of adding an orbital optimization procedure to the VQE method.

The reason behind the interest in this problem is that such a system, due to its symmetry, exhibits strong static correlation effects due to two equivalent electronic configurations arising in the bond dissociation limit. In particular, as thoroughly discussed by Olsen \textit{et al.}\cite{olsen1996full}, around the equilibrium geometry the HF determinant contribution to the FCI wavefunction largely outclasses all the other electronic configurations. On the other hand, as the bond length increases, its contribution becomes smaller and smaller until it vanishes at the dissociation limit. Simultaneously, the occupation of the orbital pair (3a$_1$, 1b$_2$), almost completely populated in the HF determinant, loses occupation to the orbital pair (4a$_1$, 2b$_2$) until they are equally populated in the dissociation limit. Here we show that, despite the inherent problem complexity, we are able to obtain a fair estimate of the solvation effects even using a variational network tailored specifically on the electronic structure of the equilibrium geometry.

\begin{figure}[h!]

\begin{tikzpicture}[node distance=cm,
    every node/.style={fill=white, font=\sffamily}]
    
    \node (figure) at (0,0) 
    {\centering
    \includegraphics[width = \textwidth]{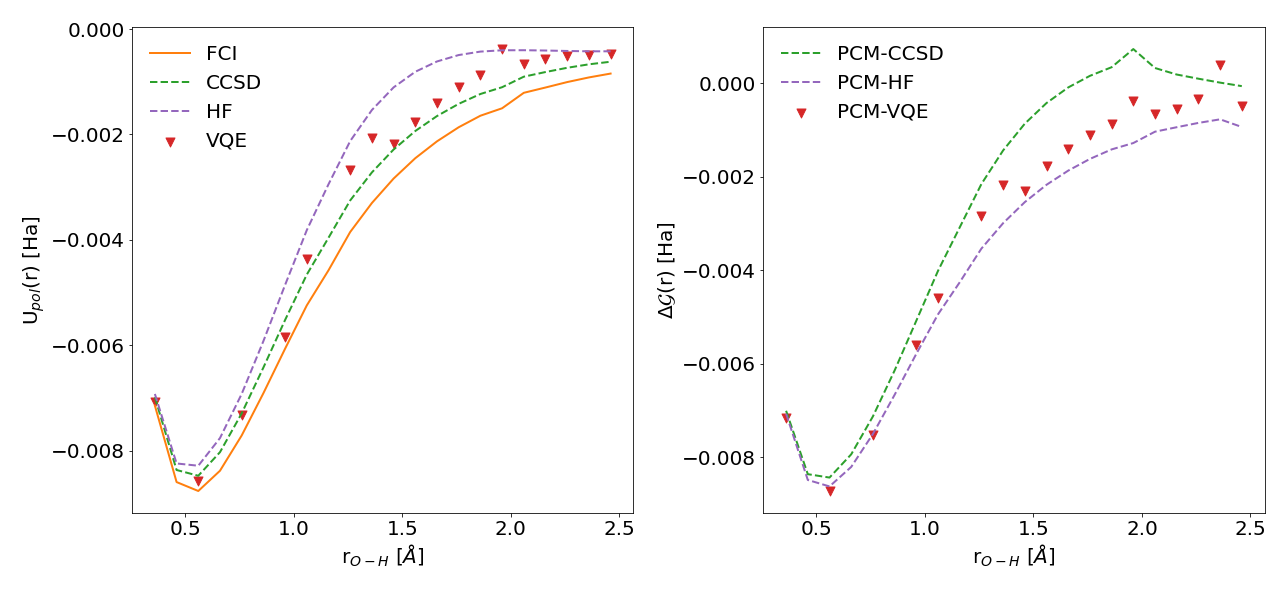}};

\node (a) at (-8.1, 3.25) {(a)};

\node (b) at (1.05, 3.25) {(b)};

\end{tikzpicture}
    \caption{Solvation effects for the water molecule (in water) along the double bond dissociation profile. (a) $U_{pol}$, polarization energy; (b) $\Delta\mathcal{G}$, solvation free energy. As a reference we report the results obtained at the FCI (orange solid line), CCSD (green dashed line) and HF (purple dashed line). Red triangles are obtained with the PCM-VQE algorithm where the number of optimization steps has been increased (up to 500) with the bond distance. For more details see SI. All calculations were performed using a STO-3G basis set.}
\label{h2o_solvation_properties}
\end{figure}

In Fig. \ref{h2o_solvation_properties}a we show the results for the polarization energy, defined as $U_{pol} = \frac{1}{2}\langle V_{\sigma} \rangle $, computed as a function of the bond distance. This quantity accounts for the interaction between the solute charge density in gas phase and the corresponding polarization charges. Due to the lack of relaxation of the electronic degrees of freedom in presence of the solvent it may deviate from the solvation free energy defined in Eq. \ref{eq:sol}.

As a first comment, we can notice that the double bond dissociation profile for the polarization energy curve predicts the greatest energetical stabilization due to the hydration process around $\approx$ 0.6\,$\AA$ which, interestingly, differs from the equilibrium distance $r_{OH} = 0.96$\,$\AA$ . For short bond lengths the PCM-VQE (red triangles) predicts values of polarization energy slightly closer to the FCI result (solid orange line) than the CCSD (green dashed line). On the other hand, as the bond length increases, we observe larger deviations from the FCI result. In spite of this, we obtain better estimates w.r.t. the HF level of theory, with a maximum variation from the CCSD estimate $<1$\,kcal/mol. It is important to emphasise that this degree of agreement is not readily apparent from the results shown in the SI regarding the absolute energy values obtained with the different methods.

Now we consider the results for the hydration free energy reported in Fig. \ref{h2o_solvation_properties}b. Here the observed qualitative behavior is reversed: the PCM-VQE predicts a slightly favourable energy contribution to the hydration process compared to CCSD for almost all bond length values considered. As noted previously, for short bond lengths, the discrepancies among the values obtained with the different methods are small; moving to the region where the strong static correlation arises larger deviations between the different approximations become more apparent with the PCM-VQE outcomes producing values in between the ones of CCSD and HF.

Finally, we want to discuss the slightly noisy trend observed in both graphs for the polarisation energy and the solvation energy estimated with the (PCM-)VQE. This could result from an interplay of different effects. First of all we highlight that the overall quality of these results could be easily further improved adopting a different ansatz such as those used in Ref. \cite{mizukami2020orbital}. Here we did not focus on the technical refinement of the implementation as was beyond the scope of the present work. Further, as shown in more detail in the SI and mentioned in the computational details, the optimisation procedure does not take place with the same number of iterations for each bond length. This choice has been dictated by the fact that, to preserve consistency with the computational procedure, in all cases the initial state for the VQE optimization was considered to be the $|$HF$\rangle$ state which is an increasingly worse ansatz as electronic correlation increases and thus requires a greater number of iterations before achieving convergence. Although the convergence rate of the optimisation procedure does not seem to be particularly affected by the presence of solvent effects in the cost functional (see SI), this may have influenced the estimation of much smaller numerical values, such as those of $\Delta\mathcal{G}$, compared to the absolute energies (or free energies). This last point is particularly relevant for r$_{\text{OH}}$ values greater than 2 $\text{\AA}$ in which the overall convergence rate is lower for both the case in vacuum and in solution w.r.t. shorter bond lengths.

In this section we have analyzed the results obtained with a PCM-VQE procedure on a noiseless quantum processor. This allows to prove the efficacy of the method but does not give us an idea of its viability on the NISQ devices that are currently available. In the following section we will look at the results obtained simulating a noisy quantum device to understand if the effects of a microscopic environment on a small molecule can be reliably caught by an actual NISQ device.

\subsection{Effect of quantum noise on solvation free energy}
\label{noise_effects_section}

The aim of this section is to understand if we are able to reliably compute with present quantum processors an estimate of the solvation free energy for a molecular system. We will focus on the six qubit system H$_3^+$ whose charged state determines a strong stabilization of the system due to the solvation process. 
Here it is important to notice, as explicitly expressed in Eq. \ref{eq:sol} that the definition of the solvation free energy involves two different sets of variational circuit parameters corresponding to the optimal result found with a standard VQE and to the optimal result after the PCM-VQE procedure.

\begin{figure}[h!]
\begin{tikzpicture}[node distance=cm,
    every node/.style={fill=white, font=\sffamily}]
\node (figure) at (0,0) {
    \centering
    \includegraphics[width =\textwidth]{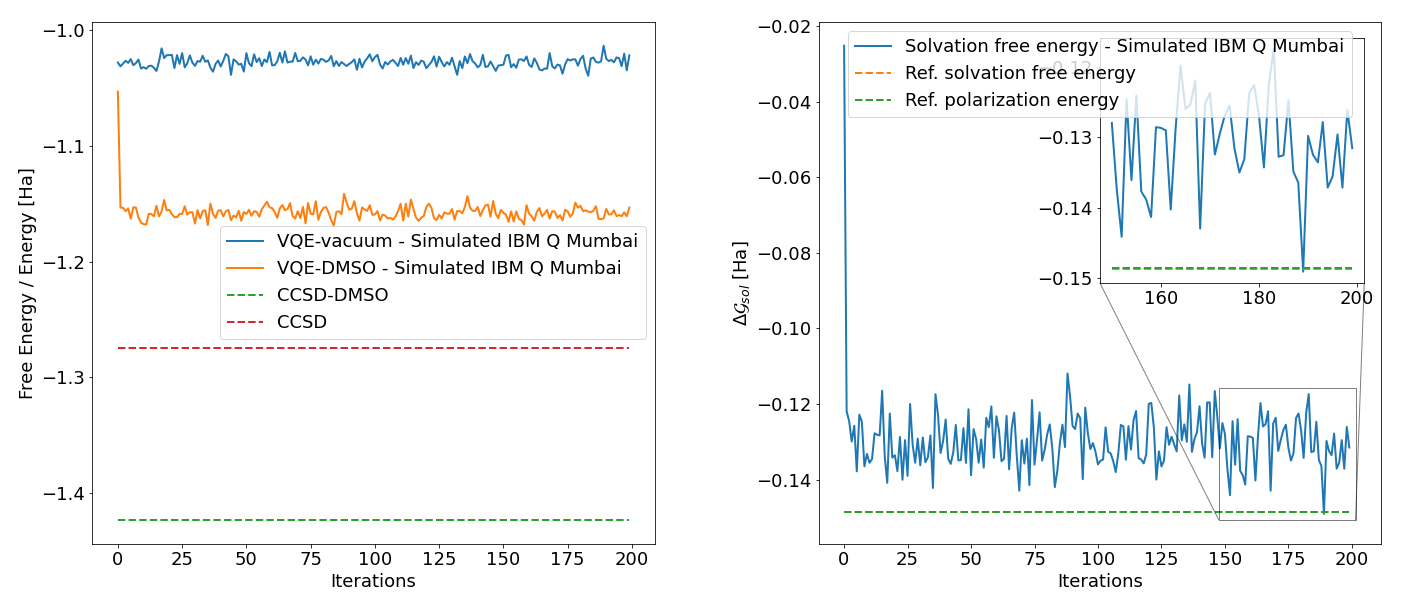}};
    
\node (a) at (-4, 3.75) {(a)};

\node (b) at (4, 3.75) {(b)};

\end{tikzpicture}

    \caption{PCM-VQE simulation of the H$_3^+$ molecule in DMSO with a noise model built upon the IBM Q Mumbai quantum processor. (a) Energy obtained with a VQE simulation in presence of the simulated quantum noise (solid blue line); Free energy in solution calculated with the PCM-VQE in presence of the simulated quantum noise (solid orange line). As a reference we show the energy obtained with a classical calculation at the CCSD level of theory (\textit{in vacuo} - red dashed line, DMSO - green dashed line). (b) Solvation free energy for the H$_3^+$ molecule in DMSO with a noise model built upon the IBM Q Mumbai quantum processor (solid blue line). As a reference we have reported the solvation free energy (orange dashed line) and polarization energy (green dashed line) computed with a CCSD calculation, these latter are superimposed and not distinguishable. The estimated value reported in the main text is obtained taking the average over the last 50 points of the iteration plot (see inset) $\pm \, \sigma$  (standard deviation of the last 50 points).}
    
\label{solvation_energy}
    
\end{figure}

In Fig. \ref{solvation_energy}a we report the free energy in solution (solid orange line) and the gas-phase energy (solid blue line) computed by running the PCM-VQE and VQE problem on a classical quantum emulator with a noise model built to mimic the IBM Q Mumbai quantum computer. The variational ansatz adopted is the same used for the calculations in noise free conditions reported in the previous section. In this case we initialized the variational parameters to random values. 

As we can see the presence of quantum noise affects the coherent state of the processor and hampers the classical optimization procedure; the result is that the in vacuo energy and free energy estimates are significantly higher w.r.t. the theoretical values reported with dashed lines in the same plot, and both continue to fluctuate with the iteration number. This is in agreement with what has been reported in other works\cite{o2016scalable, kandala2019error}. It represents the current limitations of this technology and  calls for the pressing need of error mitigation strategies. \textcolor{black}{Here we notice that the deviation (quantum vs. theoretical benchmark) on the estimated energy in vacuo and on the free energy in solution is very similar, $\Delta \approx 0.275 \, \text{Ha}$. This is reasonable since both the quantities to be measured (1,2-RDMs) and the gas-phase vs. solvated wavefunctions are very similar.} 

Indeed, as shown in Fig. \ref{solvation_energy}b, even without any error mitigation procedure, we were able to give a reasonable estimate of the solvation free energy ($\Delta \mathcal{G}_{sol}[\bar{\theta}] = \textcolor{black}{-3.564} \pm \textcolor{black}{0.163}$\,\,\textcolor{black}{eV} vs. $\Delta \mathcal{G}_{sol} = -\textcolor{black}{4.054}$\,\,\textcolor{black}{eV}) reported with the solid blue line as a function of the optimization steps. This value has been obtained performing two independent runs of the VQE algorithm \textit{in vacuo} and in solution and taking the energy difference. Further, to mitigate the effect of stochastic fluctuations, the value reported results from an average over the last 50 points of the iteration plot (see inset of Fig. \ref{solvation_energy}) and the error is estimated as $\pm \sigma$ (standard deviation of the last 50 points).

We have also analyzed the behavior of the polarization energy. As we have seen in the previous section, in general it is different from $\Delta \mathcal{G}_{sol}$ since it lacks the contribution of the wavefunction change from gas-phase to solution. However for the present system these two quantities are very similar w.r.t. the classical benchmark calculations. This is also evident in Fig. \ref{solvation_energy}b where both reference lines are completely overlapped.

We have also investigated the impact of quantum noise on the computed polarization energy. The goal of this analysis is to understand if estimating the solvation free energy approximating it with $U_{pol}$ gives better results on a NISQ device. In this case we obtained a value of $U_{pol} = \textcolor{black}{-4.054} \pm \textcolor{black}{0.217}$ \textcolor{black}{eV} that matches the quantity obtained with a noise free calculation. \textcolor{black}{Error bars are obtained according to the procedure explained in the SI and computational details' section.} 

We can read these results in light of multiple effects. First of all, as evidenced in the theory section, solvent effects are reproduced optimizing a perturbation operator whose contribution is captured by measuring only the 1-RDM. As such, the number of Pauli strings to measure is dramatically reduced. To better discuss this point we have reported the exact 1-RDM and the estimated 1-RDMs with the calculation in vacuum and in solution (see Fig. \ref{density_matrix}). Furthermore, we have computed the trace distance, \textcolor{black}{$D(\rho, \rho_{exp}) = \frac{1}{2 N_e} || \rho - \rho_{exp} ||_1$ where $N_e$ is the number of electrons, between the theoretical and experimental 1-RDMs obtaining the same value of $D = 0.075$. Note that the trace distance ranges from 0 to 1 where the former value stands for two identical RDMs and 1 is obtained for two matrices spanning two orthogonal supports.} \textcolor{black}{The same value is obtained comparing noisy vs. noiseless PCM-VQE calculations.} As we can see a quantitative agreement between the density matrices is apparent. To get the same values for the trace distance is not surprising, since the procedure to extract the 1-RDM is the same (in terms of measurement needed) both for the PCM-VQE and VQE algorithm. This is also shown by the reference values of the polarization energy and of the solvation free energy. 

On the other hand, moving to the estimated solvent effects, the outcome is soundly improved when only the polarization energy is considered. Such a result could be rationalized considering that when taking the difference between two values coming from two independent runs the errors due to the sampling of the 2-RDMs (that require a much larger number of Pauli strings to be measured) and the errors for each computed 1-RDM accumulate by propagating. Instead, when the solvation free energy is approximated with the polarization energy the errors arising from the sampling of the 2-RDM are not present. In addition we also avoid the combination of the errors coming from the subtraction of the two quantities. Since the latter are not correlated we do not expect cumulative effects but simply a propagation of more errors on the calculated quantity that contribute to worsen the accuracy of the estimated solvation free energy. \textcolor{black}{We point the reader to the SI for a more thorough analysis. We emphasise that all these results come from a combination of smaller numerical values of the matrix elements and a smaller number of Pauli words to be measured.}

Finally, it is important to stress that the accuracy w.r.t. the exact value could be further reduced applying the protocol shown in Ref.\cite{arute2020hartree} (both for $\Delta \mathcal{G}_{sol}$ and $U_{pol}$) that means post-selecting the measured values upon a total occupation number criterion and applying the McWeeny purification\cite{mcweeny1960some}. Here, since a technical optimization of the implementation goes beyond the purposes of this study, we only applied a normalization factor to the matrix elements of the 1-RDM to obtain the correct number of electrons when taking the trace.

\begin{figure}

\begin{tikzpicture}[node distance=cm,
    every node/.style={fill=white, font=\sffamily}]
    
\node (1) at (-8.5, 1) {a)};

\node (2) at (-1.25, -1.5) {d)};

\node (3) at (-1.25, 1) {b)};

\node (4) at (-8.5, -1.5) {c)};

\node (a) at (-5, 0) {$\begin{pmatrix}
    1.96 & 0.00 & 0.00 \\
0.00 & 1.85 * 10^{-2} & 0.00 \\
0.00 & 0.00 & 1.85* 10^{-2}
    \end{pmatrix}$};
\node (b) at (3, 0) {$ 
    \begin{pmatrix}
    1.67 & 2.13* 10^{-3} & -3.19* 10^{-3} \\
-2.01* 10^{-3} & 1.98* 10^{-1} & -2.60* 10^{-3} \\
1.01* 10^{-2} & -4.73* 10^{-4} & 1.35* 10^{-1} 
    \end{pmatrix}$};
\node (d) at (-5, -2) {$\begin{pmatrix}
    1.96 & 0.00 & 0.00 \\
0.00 & 1.85 * 10^{-2} & 0.00 \\
0.00 & 0.00 & 1.85* 10^{-2}
    \end{pmatrix}$};
\node (c) at (3, -2) {$ 
    \begin{pmatrix}
    1.67 & 4.60* 10^{-3} & 1.42* 10^{-3} \\
-1.53* 10^{-3} & 1.96* 10^{-1} & 5.48* 10^{-3} \\
-1.22* 10^{-2} & 7.26* 10^{-3} & 1.36* 10^{-1} 
    \end{pmatrix}$};
\end{tikzpicture}
    \caption{H$_3^+$ noisy simulation of the PCM-VQE algorithm. (a) Comparison between exact orbital 1-RDM, (b) orbital 1-RDM computed with a noisy simulation of the PCM-VQE algorithm and used to compute the solvation free energy, (c) orbital 1-RDM computed with a noise-less simulation of the PCM-VQE algorithm and (d) orbital 1-RDM computed with a noisy simulation of the VQE in gas-phase and used to compute the polarization energy.}
    \label{density_matrix}
\end{figure}


In conclusion, although the use of variational quantum algorithms for the simulation of chemical systems of interest is still hampered by the present NISQ computer accuracy, these results suggest that the technical gap to be overcome to accurately evaluate solvation effects may be lower than that to obtain accurate values of the total electronic energy.

\section{Discussion}

In this work we have proposed a direct method, without the need of additional quantum resources, to extend the Variational Quantum Eigensolver algorithm to simulate solvated systems. The analysis of the numerical results obtained from quantum simulations of simple molecules suggests that computing the solvation free energy is a computational task that can be reasonably tackled with current quantum computers.

The inclusion of solvation effects through the IEF-PCM allows to describe \textit{a plethora} of microscopic environments thus extending the computational reach of current quantum computers even more than what we have showed in this work. Indeed, here we focused on the simple (yet most common) situation of an homogeneous isotropic solvent but the inclusion of more complex environments such as metal nanoparticles or liquid-liquid phase separations is straightforward and does not require any modification of the proposed quantum algorithm. 


Future investigations that can benefit from the present work regard the integration of the proposed algorithm into more elaborated computational pipelines. The first step in this direction is the application of the method presented here to quantum algorithms that perform molecular geometry optimization being another feature strongly affected by the presence of the solvent. Other examples are the problems of calculating optical responses and excited state properties in solution that give rise to a variety of photochemical and photophysical phenomena otherwise unexplorable. In this regard, we mention the very recent contribution of Lee \textit{et al.}\cite{lee2021simulation} where the authors couple classical molecular dynamics and variational quantum simulation\cite{li2017efficient} to compute the optical response of a multi-chromophoric excitonic system.

Along with the exploration of more exotic systems with NISQ devices, this work opens up to a more theoretical question related to the theory of variational hybrid algorithms: how does the variational landscape topology change when a non-linear Hamiltonian is considered? 
The importance of this question has been remarked very recently by the work of Bittel and Kliesch\cite{bittel2021training} where the authors show (not considering observables analogue to the non-linear effective Hamiltonian used in this work) that the classical task of training a variational quantum circuit is NP-Hard. Moreover, the authors find that the complexity class of the classical optimization is not inherited by the complexity of finding the ground state of a local Hamiltonian, which is known to be QMA-Hard\cite{kempe2006complexity}, but it is an intrinsic feature of the underlying optimization landscape. Given the importance of the question for the possible developments and applications of this type of algorithms, it would be surely worth to show if their finding applies also to classes of cost functionals, such as the free energy in solution considered in this work, where the variational parameters shape not only the quantum computer state but also the measured observable. 

In conclusion, to the extent that large-scale simulation of reactivity is among the long-term goals of the computational chemistry community, and that this cannot ignore the insights provided by including environmental effects, this work paves the way for quantum simulation of molecular systems in the next generation quantum processors. We believe that the development of quantum multiscale methods must be part of the second quantum revolution agenda if we want to go beyond the current computational capabilities.

\begin{acknowledgement}

D.C.  is grateful to MIUR ”Dipartimenti di Eccellenza” under the project  Nanochemistry  for  energy  and Health (NExuS) for funding the PhD grant.

\end{acknowledgement}

\section{Associated Content}

\subsection{Supporting Information}

Additional computational details, additional results for the HeH$^{+}$ molecule with a UCCSD ansatz, \textcolor{black}{calculations on H$_3^+$ with a 6-31G basis set using an adaptive ansatz, assessment of algorithmic complexity, error estimates for finite sampling}, additional results for calculations in water. 

The Supporting Information is available free of charge at \href{http://pubs.acs.org/}{http://pubs.acs.org/}.

\bibliography{achemso-demo}
\begin{tocentry}
\centering
	\includegraphics[scale = 0.34]{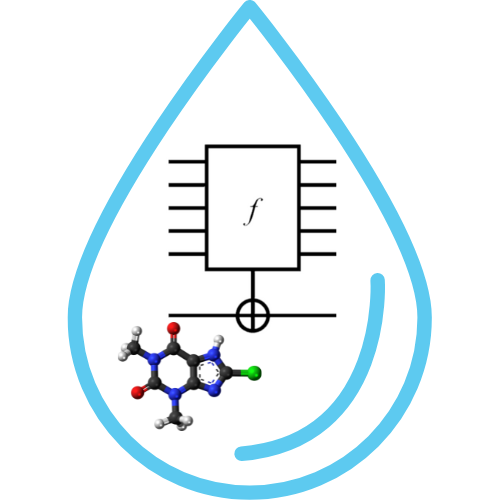}
\end{tocentry}

\end{document}